\documentclass[12pt]{article}
\usepackage{amsthm,amsfonts,amsmath}
\swapnumbers
\font\notefont=cmsl8

\theoremstyle{plain}

\newtheorem{thm}{THEOREM}[section]
\newtheorem{lm}[thm]{LEMMA}
\newtheorem{cl}[thm]{COROLLARY}

\theoremstyle{definition}

\theoremstyle{remark}
\newcommand{\upchi}{\raise1pt\hbox{$\chi$}}
\newcommand{\R}{{\mathord{\mathbb R}}}

\begin{document}

\title{\bf{The sharp constant in the Hardy-Sobolev-Maz'ya inequality in the three dimensional upper half-space}}
\author{\vspace{5pt} Rafael D. Benguria$^1$, 
Rupert L. Frank$^{2}$ and Michael Loss$^3$ \\
\vspace{-4pt}\small{$1.$ Department of Physics, P. Universidad Cat\'olica de Chile,} \\[-5pt]
\small{Casilla 306, Santiago 22, Chile, email: rbenguri@fis.puc.cl}\\
\vspace{-4pt}\small{$2.$ Matematiska institutionen, KTH Stockholm,}\\ [-5pt]
\small{100 44 Stockholm, Sweden, email: rupert@math.kth.se}\\
\vspace{-4pt}\small{$3.$ School of Mathematics, Georgia Tech,
Atlanta, GA 30332}\\ [-5pt]
\small{email: loss@math.gatech.edu}\\}
\date{May 7, 2007}
\maketitle

\footnotetext
[1]{Work partially supported by Fondecyt (CHILE) projects 106--0651 and 706--0200, 
and CONICYT/PBCT Proyecto Anillo de Investigaci\'on en Ciencia y Tecnolog\'\i a ACT30/2006. }
\footnotetext
[2]{Work partially
supported by the Swedish Foundation
for International Cooperation in Research and Higher Education (STINT).}
\footnotetext
[3]{Work partially supported by NSF-grant DMS-0600037. \\
\copyright\, 2007 by the authors. This paper may be reproduced, in its
entirety, for non-commercial purposes.}

\begin{abstract}
It is shown that the sharp constant in the Hardy-Sobolev-Maz'ya inequality on the upper half space $\mathbb{H}^3 \subset \mathbb{R}^3$
is given by the Sobolev constant. This is achieved by a duality
argument relating the problem to a Hardy-Littlewood-Sobolev type
inequality whose sharp constant is determined as well.
\end{abstract}

\section{Introduction}

The present work is concerned with a particular case of the Hardy-Sobolev-Maz'ya inequality
\begin{equation} \label{halfspace}
\int_{\mathbb{H}^n} \left[|\nabla f(\mathbf{x})|^2  -\frac{1}{4y^2}|f(\mathbf{x})|^2\right] d\mathbf{x} \ge C_n \left(\int_{\mathbb{H}^n} |f(\mathbf{x})|^{\frac{2n}{n-2}} d\mathbf{x}\right)^{\frac{n-2}{n}}
\end{equation}
where $f$ is a compactly supported function that lives in the half space
\begin{equation}
\mathbb{H}^n := \{\mathbf{x} = (x,y): x \in \mathbb{R}^{n-1} , y >0\} \ .
\end{equation}
It is quite easy to see that the left side is positive; this is
Hardy's inequality. That (\ref{halfspace}) holds for a strictly positive constant $C_n$ was proved by Maz'ya \cite{M} (Section 2.1.6., Corollary 3). 
In what follows, $C_n$ denotes the sharp constant in the above inequality.
It was shown in recent work by Tertikas and Tintarev \cite{TT}, that an optimizer
for the sharp constant $C_n$ exists provided the dimension $n \ge 4$.

The functional (\ref{halfspace}) has a number of equivalent formulations. For once it is equivalent to the inequality
\begin{equation} \label{ball}
\int_{\mathbb{B}^n} |\nabla g(\mathbf{\Omega}) |^2 d \mathbf{\Omega} 
-\int_{\mathbb{B}^n}\frac{1}{(1-|\mathbf{\Omega}|^2)^2}|g(\mathbf{\Omega})|^2 d\mathbf{\Omega} \ge C_n
 \left(\int_{\mathbb{B}^n} |g(\mathbf{\Omega})|^{\frac{2n}{n-2}} d\mathbf{\Omega} \right)^{\frac{n-2}{n}}
\end{equation}
where $\mathbb{B}^n$ is the unit ball in $\R^n$.
To see this, set
\begin{equation} \label{eff}
f(x,y)=\left(\frac{2}{(1+y)^2+ x^2}\right)^{\frac{n-2}{2}}
g(B(x,y))
\end{equation}
where $B$ is the M\"obius transformation that maps the upper half
space $\mathbb{H}^n$ to the unit ball $\mathbb{B}^n$, i.e.,
\begin{equation}
\mathbf{\Omega} = B(x,y) = \frac{\left( 2x, 1-x^2-y^2\right)}{(1+y)^2+x^2} \ .
\end{equation}
Inserting (\ref{eff}) into (\ref{halfspace}) a basically straightforward computation involving some integration by parts yields (\ref{ball}).
Clearly, this functional is invariant under rotation.
Note that these two representations, the one on the half space and the one on the unit ball show the invariance of the functional 
under all {\it M\"obius transformations} that preserve the upper half space.
This indicates that the term containing the expression $(1-|\mathbf{\Omega}|^2)^{-2}$ has some
intrinsic geometric meaning. A natural way to write the problem (\ref{halfspace}) is via stereographic projection from the unit ball
to the hyperboloid $\mathbb{P}^n$. Once more, set
\begin{equation} \label{gee}
g(\mathbf{\Omega}) = \left(\frac{2}{1-|\mathbf{\Omega}|^2}\right)^{\frac{n-2}{2}} k(P(\mathbf{u}))
\end{equation}
where
\begin{equation}
P(\mathbf{u})= \frac{\left(2\mathbf{\Omega}, 1+|\mathbf{\Omega}|^2\right)}{1-|\mathbf{\Omega}|^2} \ .
\end{equation}
It is easy to check that $P$ maps the unit ball to the upper branch
of the hyperboloid $u^2-v^2=1$, where $\mathbf{u} = (u,v)$, $u \in \mathbb{R}^n$ and $v \in \mathbb{R}$.
Inserting (\ref{gee}) into (\ref{ball}) yields the equivalent inequality
\begin{equation} \label{hyperboloid}
\int_{\mathbb{P}^n} |\nabla k(\mathbf{u})|^2 d{\rm Vol} - \frac{(n-1)^2}{4} 
\int_{\mathbb{P}^n} |k(\mathbf{u})|^2 d{\rm Vol}   \ge C_n \left(\int_{\mathbb{P}^n} |k (\mathbf{u})|^{\frac{2n}{n-2}} d{\rm Vol}\right)^{\frac{n-2}{n}} \ .
\end{equation}
The metric used here on $\mathbb{P}^n$ is the one induced by
the Euclidean space $\mathbb{R}^{n+1}$. 

As mentioned before the half space problem has been investigated in \cite{TT}, but in its formulation on 
the hyperbolic space it has also been
investigated before (see \cite{hebey} for references) although under a different point of view.
There one asks whether there exists a constant $B_n$ such that the inequality
\begin{equation}
\int_{\mathbb{P}^n} |\nabla k (\mathbf{u})|^2 d{\rm Vol} \ge S_n \left(\int_{\mathbb{P}^n} |k (\mathbf{u})|^{\frac{2n}{n-2}} d{\rm Vol}\right)^{\frac{n-2}{n}} + B_n\int_{\mathbb{P}^n} |k (\mathbf{u})|^2 d{\rm Vol}
\end{equation}
holds. Here $S_n$ is the Sobolev constant,
\begin{equation}
\frac{n(n-2)}{4} |\mathbb{S}^n|^{\frac{2}{n}}
\end{equation}
where $|\mathbb{S}^n|$ is the volume of the $n$-dimensional unit sphere in $\mathbb{R}^{n+1}$.
For $n>3$ the sharp constant  $B_n = \frac{n(n-2)}{4}$ (see \cite{hebey}). Note that $\frac{n(n-2)}{4} < \frac{(n-1)^2}{4}$.
In this language, the problem investigated in \cite{TT} is different, i.e.,  replace $B_n$ by the optimal constant and then find the sharp constant $C_n$ that will
replace $S_n$. Certainly $C_n \le S_n$, in fact $C_n < S_n$ for $n>3$.
Note that, in this case the exact value of $C_n$ is not known.

In both formulations the interesting case $n=3$ is conspicuously absent and it is this case we would like to address in this letter.  We have
\begin{thm} \label{main}
The inequality
\begin{equation}
\int_{\mathbb{H}^3} |\nabla f(\mathbf{x})|^2 d\mathbf{x} \ge \int_{\mathbb{H}^3} \frac{1}{4y^2}|f(\mathbf{x})|^2 d\mathbf{x} + S_3\left(\int_{\mathbb{H}^3} |f(\mathbf{x})|^{6} d\mathbf{x} \right)^{\frac{1}{3}}
\end{equation}
holds where $S_3$ is the sharp Sobolev constant in three dimensions, i.e.,
\begin{equation}
S_3 = 3({\pi /2})^{4/3} \ .
\end{equation}
The inequality is always strict for nonzero $f$'s. Using the formulation on hyperbolic space we have the inequality
\begin{equation}
\int_{\mathbb{P}^3} |\nabla k (\mathbf{u})|^2 d{\rm Vol} \ge S_3 \left(\int_{\mathbb{P}^3} |k (\mathbf{u})|^{6} d{\rm Vol}\right)^{\frac{1}{3}} + \int_{\mathbb{P}^3} |k (\mathbf{u})|^2 d{\rm Vol} \ .
\end{equation}

\end{thm}

In contrast to the case $n=3$, for $n \ge 4$ the sharp constant is always attained for some nonzero function (see \cite{TT}).

The problem (\ref{halfspace}) has been generalized to the case where
the underlying domain $D$ is a convex set. In this case one replaces
$\frac{1}{4y^2}$ by $\frac{1}{4d(x)^2}$ where $d(x)$ is the distance
of the point $x \in D$ to the boundary of $D$. 
It is conjectured in \cite{TT} that the
sharp constant for convex domains is given by the half space problem.
This is true for the case where the domain is a ball. We have
\begin{thm} \label{ballinequ}
The inequality
\begin{equation}\label{inequonball}
\int_{\mathbb{B}^n} |\nabla g(\mathbf{\Omega}) |^2 d\mathbf{\Omega} 
-\int_{\mathbb{B}^n}\frac{1}{4(1-|\mathbf{\Omega}|)^2}|g(\mathbf{\Omega})|^2 d\mathbf{\Omega} \ge C_n \left(\int_{\mathbb{B}^n} |g(\mathbf{\Omega})|^{\frac{2n}{n-2}} d\mathbf{\Omega} \right)^{\frac{n-2}{n}}
\end{equation}
holds for all 
smooth functions compactly supported in the unit ball. For nonzero $g$'s the inequality is always strict. 
\end{thm}
The inequality follows directly from (\ref{ball}) by noting that for
$|\mathbf{\Omega}| < 1$,
\begin{equation}
\frac{1}{(1-|\mathbf{\Omega}|^2)^2} > \frac{1}{4(1-|\mathbf{\Omega}|)^2} \ .
\end{equation}
That the inequality is sharp and always strict for non-zero functions
can be seen by scaling down a compactly supported `almost' optimizer of the half space problem and use this as a trial function
for the ball problem. Note that this device also works for general convex domains. The hard part is to establish the analog of (\ref{inequonball}) for general convex domains.

An amusing consequence of the formulation (\ref{ball}) is that
by inversion with respect to the unit sphere one obtains a sharp inequality on the complement of the unit ball, i.e., we have
\begin{thm} \label{annulus}  
The inequality
\begin{equation}
\int_{(\mathbb{B}^n)^c} |\nabla g(\mathbf{\Omega}) |^2 d\mathbf{\Omega} 
-\int_{(\mathbb{B}^n)^c}\frac{1}{(1-|\mathbf{\Omega}|^2)^2}|g(\mathbf{\Omega})|^2 d\mathbf{\Omega} \ge 
C_n \left(\int_{(\mathbb{B}^n)^c} |g(\mathbf{\Omega})|^{\frac{2n}{n-2}} d\mathbf{\Omega} \right)^{\frac{n-2}{n}}
\end{equation}
holds for all functions that are smooth and have compact support on
$(\mathbb{B}^n)^c$ the complement of the ball $\mathbb{B}^n$ in $\R^n$. Moreover, for $n> 3$ equality can be attained in the sense
of \cite{TT}.
\end{thm}
The appropriate formulation of this inequality for general domains, not necessarily convex, is an open problem. Theorem \ref{annulus}
suggests that the `correct' inequality is formulated in terms
of either the {\it harmonic radius} or the {\it hyperbolic radius}
of a domain $D$. For a definition of these concepts we refer the reader to \cite{bandleflucher}. Both of these objects are conformally
covariant, i.e., under conformal transformations they scale with
the $n$-th root of the Jacobian. In the case of a ball, the two concepts coincide and are equal to $(1-|\mathbf{\Omega}|^2)$.
Since the ball and the half space are conformally the same, these
two concepts coincide also on the half space and are given by
$2y$. Thus, it is natural to ask for which domain $D$ does the inequality
\begin{equation}
\int_D \left[|\nabla f|^2  - \frac{1}{R(x)^2} |f(x)|^2\right]d^nx
\ge C_n \left( \int_D |f(x)|^{\frac{2n}{n-2}} d^n x \right)^{\frac{n-2}{n}}
\end{equation}
hold. Here $R(x)$ is either the harmonic radius or the hyperbolic radius. In this formulation, due to its conformal invariance,
one might be able to show that the Hardy-Sobolev-Maz'ya inequality for general convex domains holds with the same constant as the one on the half space.

The plan of the paper is the following. In Section \ref{greenfct}
we derive the Green function for fractional powers of the operator
$-\Delta - \frac{1}{4y^2}$. 
This yields Hardy-Littlewood-Sobolev
type kernels. In Section \ref{hlsmazya} we prove $L^p$ estimates
for these kernels and recover Theorem \ref{main}.

\section{The Green function}\label{greenfct}
It is convenient to start with the following heat type equation on the
upper half space $\mathbb{H}^n$ 
\begin{equation}\label{initial}
u_t = \Delta u + \frac{1}{4y^2} u  \ , u(x,y;0)=f(x,y) \ .
\end{equation}
Substituting $u=\sqrt y g$ one obtains the equation
\begin{equation}
g_t = \Delta_x g + g_{yy} + \frac{1}{y} g_y \ , g(x,y;0)= \frac{f(x,y)}{\sqrt y} \ ,
\end{equation}
and one see that the right side of the equation is an $n+1$ dimensional Laplacian.
Note that $g_{yy} + \frac{1}{y} g_y$ is the two dimensional Laplacian
of a radial function.
A similar idea has been used in \cite{Beckner} in a different context.
With this in mind one arrives at once at the following formula for the solution of the heat equation
\begin{equation}
u(x,y;t) = \int_{\mathbb{H}^n} G(x-x',y,y';t) f(x',y') dx'dy'
\end{equation}
where  
\begin{equation}
G(x-x',y,y';t) = \left(\frac{1}{4\pi t}\right)^{\frac{n+1}{2}} \sqrt {y y'} e^{-\frac{(x-x')^2 +y^2+y'^2}{4t}} \int_0^{2\pi} e^{ \frac{yy'}{2t}\cos \phi} d \phi  \ .
\end{equation} 
It is not hard to see that this heat kernel is a contraction semigroup
on $L^2(\mathbb{H}^n)$ with Lebesgue measure. Thus, the generator
$Q$ is a selfadjoint operator and it is an extension of $-\Delta -\frac{1}{4y^2}$ originally defined on smooth functions with compact support in $\mathbb{H}^n$. Note that the $L^2$-norm of the gradient of functions in the domain of $Q$ is in general not finite. We shall continue to use the symbol $-\Delta -\frac{1}{4y^2}$ to denote $Q$.

It is straight forward to see (see e.g., Theorem 7.10 in \cite{Lieb-Loss}) that
\begin{equation}
\lim_{t \to 0}\frac{1}{t} \left[\Vert f \Vert^2_{L^2(\mathbb{H}^n)} - (f, G_t f)_{L^2(\mathbb{H}^n)}\right]
= 2 \pi \int_{\mathbb{H}^n} \left( |\nabla_x g|^2+|g_y|^2\right) y dy dx
\end{equation}
where $G_t f$ is the solution of the intial value problem (\ref{initial}) and $g=\frac{f}{\sqrt y}$. Note that the right hand side is manifestly positive and coincides with the interpretation of $ -\Delta -\frac{1}{4y^2}$ given in \cite{TT}.

Via the heat kernel it is straightforward to find the kernel of the fractional powers
\begin{equation}
(-\Delta-\frac{1}{4y^2})^{-\frac{\alpha}{2}}(\mathbf{x};\mathbf{x'}) = \frac{1}{\Gamma(\frac{\alpha}{2})}
\int_0^\infty  t^{\frac{\alpha}{2}} G(x-x',y,y';t) \frac{dt}{t} \ ,
\end{equation}
for $\alpha > 0$, and a calculation leads to the expression
\begin{eqnarray}
& &(-\Delta-\frac{1}{4y^2})^{-\frac{\alpha}{2}}(\mathbf{x};\mathbf{x'}) \\
&=&2^{-\alpha} \pi^{-\frac{n+1}{2}} \frac{\Gamma(\frac{n+1-\alpha}{2})}{\Gamma(\frac{\alpha}{2})} \sqrt {y y'}
\int_0^{2\pi} \left[(x-x')^2 + y^2+y'^2 - 2yy'\cos \phi \right]^{-\frac{n+1-\alpha}{2}} d \phi \\
&=:& \Phi_{n,\alpha}(\mathbf{x};\mathbf{x'})  \ .
\end{eqnarray}
Similarly, well known expressions hold for 
$(-\Delta)^{-\frac{\alpha}{2}}$
on $\R^n$ which, for reasons that become clear later, we write
in terms of the variables $(x,y)$ as
\begin{eqnarray}
& &(-\Delta)^{-\frac{\alpha}{2}}(\mathbf{x};\mathbf{x'})\\
&=&2^{-\alpha} \pi^{-\frac{n}{2}}  \frac{\Gamma(\frac{n-\alpha}{2})}{\Gamma(\frac{\alpha}{2})} \left[(x-x')^2+(y-y')^2\right]^{-\frac{n-\alpha}{2}} \\
&=:& \Psi_{n,\alpha}(\mathbf{x};\mathbf{x'}) \ .
\end{eqnarray}

First we state some simple pointwise properties about the kernel $\Phi_{n,\alpha}$.

\begin{lm} \label{pointwise} If $ n \le \alpha \le n+1$, we have that
\begin{equation}
\sup_a \Phi_{n,\alpha}(x,y+a; x',y'+a)=
\lim_{a \to \infty}\Phi_{n,\alpha}(x,y+a; x',y'+a) \equiv \infty \ .
\end{equation}
If $ n-1 \le \alpha < n $
we have that
\begin{equation}
\sup_a \Phi_{n,\alpha}(x,y+a; x',y'+a) =\lim_{a \to \infty}\Phi_{n,\alpha}(x,y+a; x',y'+a) \equiv \Psi_{n,\alpha}(\mathbf{x};\mathbf{x'}) \ .
\end{equation}
\end{lm}
\begin{proof}
An elementary calculation shows that
\begin{equation}
 \Phi_{n,\alpha}(\mathbf{x};\mathbf{x'}) = |\mathbf{x}-\mathbf{x'}|^{-n+\alpha}
2^{-\alpha} \pi^{-\frac{n+1}{2}} \frac{\Gamma(\frac{n+1-\alpha}{2})}{\Gamma(\frac{\alpha}{2})} F(A)
  \ ,
 \end{equation}
 where 
 \begin{equation}
 A= \frac{\sqrt{yy'}}{|\mathbf{x}-\mathbf{x'}|}
 \end{equation}
 and
 \begin{equation}
 F(A) :=\int_{-\pi }^{\pi } \frac{A}{\left[1+2A^2(1-\cos(\phi))\right]^{\frac{n+1-\alpha}{2}}} d\phi \ .
 \end{equation}
All the statements are an immediate consequence of Lemma \ref{estimates} with $\beta = \frac{n+1-\alpha}{2}$.
\end{proof}

\section{$L^p$-estimates for fractional powers} \label{hlsmazya}
As a consequence of Lemma \ref{pointwise} and Lieb's sharp constant in the Hardy-Littlewood Sobolev inequality \cite{Lieb} we have the following corollary.
\begin{cl} \label{HHLS} If $ n \le \alpha \le n+1$ then the operator
\begin{equation}
(-\Delta-\frac{1}{4y^2})^{-\frac{\alpha}{2}}
\end{equation}
is not bounded on $L^p(\mathbb{H}^n)$ for any $1 \le p \le \infty$.
If $n-1 \le \alpha < n$ then this operator
is a bounded operator from $L^p(\mathbb{H}^n)$ to $L^q(\mathbb{H}^n)$
for all $1< p,q < \infty$ that satisfy
\begin{equation}
\frac{1}{q} = \frac{1}{p} -\frac{\alpha}{n} \ .
\end{equation}
Moreover, for such values of $\alpha$ we have
\begin{equation} \label{hardylittlesob}
(f, (-\Delta-\frac{1}{4y^2})^{-\frac{\alpha}{2}}f)
\le 2^{-\alpha} \pi^{-\frac{n}{2}}  \frac{\Gamma(\frac{n-\alpha}{2})}{\Gamma(\frac{\alpha}{2})}C(n,\alpha) \Vert f \Vert_p^2
\end{equation}
where $p= \frac{2n}{n+\alpha}$ and
\begin{equation}
C(n,\alpha) = \pi^{\frac{n-\alpha}{2}}
\frac{\Gamma(\frac{\alpha}{2})}{\Gamma(\frac{n+\alpha}{2})}
\left[\frac{\Gamma(\frac{n}{2})}{\Gamma(n)}\right]^{-\frac{\alpha}{n}}
\end{equation}
is the sharp constant. This constant is not attained in 
(\ref{hardylittlesob}) for nonzero functions.
\end{cl}

\begin{proof}[Proof of Theorem \ref{main}]
We write
\begin{equation}
|(f,g)| = |(Q^{\alpha/4}f, Q^{-\alpha/4}g)|
\le (f, Q^{\alpha/2}f)^{1/2} (g, Q^{-\alpha/2}g)^{1/2}
\end{equation}
which by Corollary \ref{HHLS} yields the bound
\begin{equation} 
|(f,g)|^2 \le 2^{-\alpha} \pi^{-\frac{n}{2}}  \frac{\Gamma(\frac{n-\alpha}{2})}{\Gamma(\frac{\alpha}{2})}C(n,\alpha) (f, Q^{\alpha/2}f) \Vert g \Vert_p^2
\end{equation}
for $n-1 \le \alpha < n$ and $p =  \frac{2n}{n+\alpha}$. Thus,
\begin{equation}
\Vert f \Vert_{p'}^2 <  2^{-\alpha} \pi^{-\frac{n}{2}}  \frac{\Gamma(\frac{n-\alpha}{2})}{\Gamma(\frac{\alpha}{2})}C(n,\alpha)(f, Q^{\alpha/2}f) \ ,
\end{equation}
and there is never equality in the above inequality for nonzero functions. Theorem \ref{main} follows by choosing $n=3$ and $\alpha=2$.
\end{proof}

The reader may wonder what happens when $0< \alpha <n-1$. While we do not succeed in calculating the sharp constant, it is possible to
show that the sharp constant in inequality (\ref{hardylittlesob}) is attained. While this constant is strictly
bigger than the corresponding constant in the Hardy-Littlewood-Sobolev
inequality, we do not know its exact value.

The procedure for proving this relies on the conformal invariance 
of the kernel which allows to transform the problem into
one on the unit ball.
Then the device of competing symmetries developed in
\cite{Carlen-Loss} allows to restrict the maximization problem to radial functions on the ball. The correction term to Fatou's lemma
(\cite{Brezis-Lieb}, see also \cite{Lieb-Loss}) then allows to show the existence
of a maximizer. Thus, we recover some of the results in \cite{TT} with
a different proof. Moreover, it is also possible to show that every maximizer
is the conformal image of a radial function. The details will
appear elsewhere.

\section{Appendix }
In this appendix we collect some facts about the function
\begin{equation}
F(A) := \int_{-\pi}^\pi \frac{A}{(1+2A^2(1-\cos(\phi))^\beta} d\phi \ ,
\end{equation}
where $\beta = \frac{n+1-\alpha}{2}$.
\begin{lm} \label{estimates}
Depending on the value of $\beta$, the function $F(A)$ has the following asymptotics as $A \to \infty$.
\item{a)} If $0 \le \beta \le \frac{1}{2}$ then
$\lim_{A \to \infty} F(A) = \infty$.
\item{b)} If $\frac{1}{2} < \beta \le 1$, then $F(A)$ is a monotone increasing function and
\begin{equation}
\lim_{A \to \infty} F(A) = \sqrt{\pi} \frac{\Gamma(\beta -\frac{1}{2})}{\Gamma(\beta)} \ .
\end{equation}
\end{lm}
\begin{proof}
Since
\begin{equation}
F(A)= \int_{-\pi A}^{\pi A} \frac{1}{1+2A^2(1-\cos(\frac{\phi}{A}))^\beta} d \phi
\end{equation}
the limit as $A \to \infty$ is
\begin{equation}
\int_{-\infty}^\infty \frac{1}{(1+\phi^2)^\beta} d\phi = 
\sqrt{\pi} \frac{\Gamma(\beta -\frac{1}{2})}{\Gamma(\beta)}
\end{equation}
for $\beta > \frac{1}{2}$ and it is $+\infty$ for $\beta \le \frac{1}{2}$. This proves a). To see that b) holds for $\beta=1$
one easily performs the $\phi$ integration and obtains
\begin{equation}
F(A)= \frac{2\pi A}{\sqrt{1+4A^2}} \ ,
\end{equation}
which is obviously increasing with $A$.
For $\frac{1}{2} < \beta < 1$ we use the formula
\begin{eqnarray}
& &\left[1+2A^2(1-\cos \phi) \right]^{-\beta} \\
&=&\frac{\sin(\pi \beta)}{\pi} \int_0^\infty \left[1+t+2A^2(1-\cos \phi) \right]^{-1} t^{1-\beta} \frac{dt}{t} \ .
\end{eqnarray} 
Integrating with respect to $\phi$ yields
\begin{equation}
F(A) = 2\sin(\pi \beta) \int_0^\infty \frac{A}{\sqrt{ (1+t)^2 +4(1+t)A^2}} t^{1-\beta}\frac{dt}{t} \ .
\end{equation}
Again, this function increases with $A$. 
\end{proof}

\end{document}